\documentstyle[12pt]{article}
\newcommand{\be}{\begin{equation}}
\newcommand{\ee}{\end{equation}}
\def\n{\noindent}

\catcode `\@=11
\@addtoreset{equation}{section}
 
\catcode `\@=12

\title{\bf\huge An ansatz for spacetimes of zero gravitational mass : \\                                     global monopoles and textures}     
\author{Naresh Dadhich\thanks{E-mail : nkd@iucaa.ernet.in} \\
{\sl Inter-University Centre for Astronomy \& Astrophysics,}\\
{\sl Post Bag 4, Ganeshkhind, Pune - 411 007, India.} \\
K. Narayan \thanks{E-mail :  narayan@niharika.phy.iitb.ernet.in}
{\sl Department of Physics,}\\
{\sl Indian Institute of Technology Bombay,} \\
{\sl Mumbai 400 076.}
} 
\date{}
\begin{document}
\maketitle

\begin{abstract}
We    propose    a   geometric   ansatz,   a    restriction    on 
Euclidean / Minkowski   distance  in  the  embedding  space   being 
propotional  to  distance  in the  embedded  space,  to  generate 
spacetimes with vanishing gravitational mass ($R_{ik} u^i u^k = 0,
u_i u^i = 1 $). It 
turns  out that these spacetimes can represent  global  monopoles 
and  textures. Thus the ansatz is a prescription to generate  zero 
mass  spacetimes that could describe topological defects,  global 
monopoles and textures. 
\end{abstract}


\vfill
\begin{flushright} IUCAA-32/97 \end{flushright}
\vfill

\newpage

\n By  the restriction that  Euclidean / Minkowski distance  in  the 
embedding  space is proportional to the distance in the  embedded 
space  we  obtain  spacetimes with  the  vanishing  gravitational 
charge  density  defined by, $- 4 \pi \rho_c = R_{ik} u^i u^k =
 - 4 \pi (\rho + 3p), u_i u^i = 1$. That 
means active gravitational mass of the spacetime is zero. This is 
also   the  characteristic  property  of  spacetimes   describing 
topological defects like global monopoles and global textures [1-
4].   Such  exotic objects are supposed to be  created  by  phase 
transition  in  the  early  Universe  when  global  symmetry   is 
spontaneously  broken.  In  recent times  there  have  been  some 
interesting  applications of such objects in the  early  Universe 
cosmology [1-11].  \\

\n Newtonian theory (NT) is a linear scalar theory, in which absence 
of  gravitational  field  is characterised by  constancy  of  the 
scalar  potential, $\phi$. The physical measurable effects of  gravity 
are particle and tidal accelerations.For  the canonical situation 
of  field due to a mass point, the former is always radial  while 
the  latter is also experienced only by radially  falling  nearby 
particles. They are given by first and second  derivatives of 
potential  relative to radial coordinate $r$ and hence both  vanish 
when $\phi = const.$ \\

\n Let us consider this question in general relativity (GR). How
does one define potential in GR ? In the context of the Schwarzschild
field, we would like to make the following working definition. By
potential we shall mean a scalar function that (i) completely describes
the field, (ii) is a solution of the Laplace equation corresponding to
$R^t_t = 0$, (iii) defines equivalent Newtonian potential and (iv) determines
radial acceleration for a free particle. Since GR is a non-linear theory,
this scalar function also appears in the Riemann curvature as it is, 
which will in contrast
to NT impart a physical meaning to it. This is the new and unique feature
of GR, which does not consonant with the usual concept of potential in
classical physics. We shall in what follows mean by it as stated above,
essentially a function that completely characterises the Schwarzschild
field and shall call it relativistic potential. \\ 

\n In GR, gravitation is described by 
curvature  of spacetime. Relativistic potential 
not only gives rise  to  the   
acceleration  but it also ``curves'' the space part 
of  the  metric. Note that we could have  acceleration  
non-zero with space part of the metric being flat. It can  be 
shown that non-linear feature of gravitation (field being its own 
source)  goes  into curving the space part of the  metric,  which 
survives  even when potential is constant but not  zero  [12-14]. 
This  is  how  constant  potential in GR  gives  rise  to  curved 
spacetime. Of  course this will have zero gravitational  mass  and 
free  of  radial acceleration as well as tidal  acceleration  for 
radial  motion. The curvature shows up only in tidal  acceleration 
for  transverse  motion, which was however absent in  NT  because 
space  was  flat.  This spacetime was  first  considered  by  the 
author [15]  as an example of a spacetime, having of 20  curvatures 
all   but   one  non-zero. It  gave  rise   to   a   stress-system                
$T^t_t = T^r_r \sim 1/r^2$, and  the rest of stresses being zero. 
This did not accord  to  any 
conventional matter fields and hence was not pursued any  further 
and  forgotten. On the other hand this is precisely  the  stresses 
required to describe a global monopole [1]. It is remarkable  that 
the  spacetime  arising from a constant relativistic potential  can  actually 
describe an exotic object like global monopole, which is supposed 
to  be  created  by spontaneously  breaking  of  global  symmetry        
$O(3)$ into $U(1).$ \\

\n The  simplest  model of  a global monopole is  described  by  the 
Lagrangian [1],

\begin{equation}
L = \frac{1}{2} \partial_i \psi^a \partial^i \psi^a - \frac{\lambda}{4}
(\psi^a \psi^a - \eta^2)^2 
\end{equation}

\n where $\psi^a$   is  a triplet of scalar fields, $\psi^a = \eta f(r)
x^a/r$. It  has  global 
$O(3)$ symmetry   which  is   spontaneously   broken   to 
$U(1)$. Outside the monopole core $f = 1$, and $\psi^a$ takes the vacuum value      $\eta x^a/r$ and stresses will be given by

\begin{equation}
T^t_t = T^r_r = \frac{\eta^2}{r^2}
\end{equation}

\n and rest of $T^k_i = 0.$ \\

\n Let us consider the spherically symmetric metric,

\begin{equation}
ds^2 = A dt^2 -Bdr^2 - r^2 d \Omega^2, d \Omega^2 = d \theta^2 + \sin^2 \theta
d \varphi^2 .
\end{equation} 

\n Now $T_{tr} = 0$    and $T^t_t = T^r_r$    lead to $A = B^{-1} = 1 + 2 \phi(r)$   and the two equations 

\begin{equation}
R^t_t = - \bigtriangledown^2 \phi = 0
\end{equation}

\begin{equation}
R^{\theta}_{\theta} = R^{\varphi}_{\varphi} = - \frac{2}{r^2} (r \phi)^{\prime}
\end{equation}

\n where a dash denotes derivative w.r.t. $r$ . \\ 

\n The  former is the good old Laplace equation admitting the  well-
known  general solution $\phi = K - M/r$ 
while the latter determines  the 
stresses (2) when $K = -4 \pi \eta^2 \neq 0$.
Thus the Schwarzschild  black  hole 
with global monopole charge is described by the metric [1], 

\begin{equation}
ds^2 =(1 - 8 \pi \eta^2 - \frac{2M}{r}) dt^2 - (1 - 8 \pi \eta^2
- \frac{2M}{r})^{-1} dr^2 - r^2 d \Omega^2 .
\end{equation}

\n Note  that $\phi = K \neq 0$  gives rise to  curved  spacetime  representing 
stresses  corresponding to a global monopole. It goes over to  the 
Schwarzschild field when $K = 0.$ \\
  
\n In this note we shall propose an ansatz to obtain spacetimes with 
vanishing gravitational charge density (zero mass). Let us  begin 
with a 5-dimensional flat spacetime 

\begin{equation}
ds^2 = dt^2 - dx^2_1 - dx^2_2 - dx^2_3 - dx^2_4
\end{equation}

\n and impose the restriction 

\begin{equation}
x^2_1 + x^2_2 + x^2_3 + x^2_4 = k^2 (x^2_1 + x^2_2 + x^2_3)
\end{equation}

\n where $k$ is a constant. Now elimination of $x_4$  will lead to the global 
monopole metric, 

\begin{equation}
ds^2 = dt^2 - k^2 dr^2 - r^2 d \Omega^2
\end{equation}

\n which  is  equivalent  to  (6)  with $k^2 = (1 - 8\pi \eta^2)^{-1}$
and $M = 0$. Here  $t$  has   not 
participated in the restriction (8). \\
 
\n If we had instead considered the restriction 

\begin{equation}
x^2_2 + x^2_3 + x^2_4 = k^2 (x^2_2 + x^2_3)
\end{equation}

\n which  would  have  led to the metric for a  cosmic  string  with 
deficit angle and vanishing Riemann curvature. \\
 
\n If we let $t$ also in and consider

\begin{equation}
t^2 - x^2_1 - x^2_4 = k^2(t^2 - x^2_1)
\end{equation} 

\n which would lead to the metric 

\begin{equation}
ds^2 = k^2 d \mu^2 - \mu^2 d \theta^2 - d x^2_2 - d x^2_3
\end{equation}

\n where $\mu^2 = t^2 - x^2_1$. 
This  is  also  a flat  spacetime  with  a  defecit 
angle.It may be considered as Lorentzian version of cosmic string 
for $\mu^2 = t^2 - x^2_1$. In general a metric with a defecit angle can be obtained by letting any two of the variables participate in the ansatz. \\
 
\n Analogus to Lorentzian cosmic string, let us consider  Lorentzian 
monopole by 

\begin{equation}
t^2 - x^2_1 - x^2_2 - x^2_4 = k^2 (t^2 - x^2_1 - x^2_2)
\end{equation}

\n which will yield the metric, 

\begin{equation}
ds^2 = k^2 d \mu^2 - \mu^2 d \theta^2 - \mu^2 \sin h^2 \theta
d \varphi^2 - d x^2_3
\end{equation}

\n where $\mu^2 = t^2 - x^2_1 - x^2_2$. It gives rise to 

\begin{equation}
T^{\mu}_{\mu} = T^{x_3}_{x_3} = \frac{1-k^2}{8 \pi k^2\mu^2},
R^{\theta \varphi}_{~~\theta \varphi} = \frac{1-k^2}{k^2 \mu^2}
\end{equation}

\n and other components vanishing. \\
 
\n This  is  similar  to the monopole  case. These  stresses  can  be 
formally  generated  by  the same Lagrangian  (1)  with  $r$  being 
replaced  by $\mu$  and $T^{\mu}_{\mu} = T^{x_3}_{x_3} = 
\eta^2/\mu^2 $. This can be  looked  upon  as   
Lorentzian version of global monopole. \\

\n Finally if we let all the variables to participate in the ansatz, 

\begin{equation}
t^2 - x^2_1 - x^2_2 - x^2_3 - x^2_4 = k^2 (t^2 - x^2_1 - x^2_2 - x^2_3)
\end{equation}

\n which will give the metric 

\begin{equation}
ds^2 = k^2 d \mu^2 - \mu^2 d \chi^2 - \mu^2 \sin h^2 \chi d \Omega^2
\end{equation}

\n with $\mu^2 = t^2 - x^2_1 - x^2_2 - x^2_3$.
This will give rise to perfect fluid with  the 
equation of state $\rho + 3p = 0$,

\begin{equation}
8 \pi \rho = \frac{3(1-k^2)}{k^2 \mu^2}, R^{\theta \varphi}_{~~\theta
\varphi} = R^{\theta \chi}_{~~\theta \chi} = R^{\varphi \chi}_{~~\varphi \chi}
= \frac{1- k^2}{k^2 \mu^2}.
\end{equation}

\n This will correspond to a spacetime of global texture  [2,10].Again 
the  above  stresses  can formally be  generated  from  the  same 
Lagrangian  (1) with $r$ being replaced by the Lorentzian $\mu$  and  we 
get  

\begin{equation}
T^{\mu}_{\mu} = \frac{3 \eta^2}{2 \mu^2}, T^{\chi}_{\chi} = T^{\theta}_{\theta}
= T^{\varphi}_{\varphi} = \frac{\eta^2}{2 \mu^2}. 
\end{equation}

\n This  is  the equation of state $\rho + 3p = 0$ 
employed to  characterise  vacuum 
energy  or  a  new  kind  of  exotic  matter  called  k-matter[5-
8]. Several  authors have considered cosmological implications  of 
decaying vacuum energy and time varying cosmological term[5-8, 16-
17, and others]. Here we have given an interesting and novel way of 
generating  this  kind  of  exotic  vacuum  energy  distribution, 
characterised  by the equation of state $\rho + 3p = 0$. This  indicates 
that spacetime has zero active gravitational mass. \\
                 
\n Thus we have given an elegant geometric ansatz for generation  of 
spacetimes  of  global  monopoles  and  textures  with  vanishing 
gravitational  mass. The  ansatz is simply the  restriction,  that 
``distance'' in the embedding space is propotional to ``distance'' in the 
embedded  space. This  generates curvature in  embedded  spacetime 
corresponding  to vanishing gravitational mass. This is  also  the 
characteristic property of topological defects, global  monopoles 
and  textures. Here we do formally obtain  Lorentzian versions  of 
cosmic  strings and global monopoles. Such objects have  not  been 
considered  in literature. Their physical interpretation
 is an open issue. They  formally follow from the ansatz 
when Euclidean distance is replaced by Lorentzian distance in the 
field  configration. Global texture is on the other hand is a   4-
dimentional   generalization   of  global  monopole. It   can   be 
considered as generalization of Lorentzian monopole as well. \\
 
\n It  is obvious from the ansatz that all these spacetimes will  be 
by  definition  embeddable in 5-dimentional  Minkowski  spacetime 
(i.e. they are of class I). What is perhaps not so obvious is  the 
fact  that they are not conformally flat. 
It would be interesting  to  consider 
spacetimes  conformal to these zero mass spacetimes  and  examine 
whether  they  represent any specific properties. One  of  us (ND) has 
studied spacetimes conformal to global monopole metric (9) and has 
shown  that it picks up uniquely isothermal (density falling  off 
as  inverse square of the area radius with a linear  equation  of 
state) character of fluid sphere without boundary. It is a general 
result   that  a  necessary  and  sufficient  condition  for   an 
isothermal  fluid sphere without boundary is that its  metric  is 
conformal  to  the monopole metric (9) [18-19]. We  shall  consider 
separately spacetimes conformal to Lorentzian global monopole and 
texture spacetimes [20]. \\

\n Global texture metric (17) is clearly not conformally flat instead
it is conformal to hyperbolic static (Einstein) metric with negative 
curvature. There do however exist conformally
flat spacetime with the equation of state $\rho + 3p = 0$. It is given
by

\begin{equation}
ds^2 = e^{2 \alpha t} (dt^2 - dr^2 - r^2 d \Omega^2), \alpha = const.
\end{equation}

\n while the global texture will have the form

\begin{equation}
ds^2 = e^{2 \eta/k} (d \eta^2 - \frac{d R^2}{1 + R^2} - R^2 d \Omega^2).
\end{equation}

\n The two are different spacetimes because the base metric for (20)
is flat while for (21) it is hyperbolic Einstein spacetime. In the latter
case, both base as well as conformal spacetimes obey the equation of
state $\rho + 3p = 0.$ \\

\n Since  gravitational mass $R_{ik} u^i u^k = 0$
is zero, the relativistic potential is constant, making acceleration
as well as tidal acceleration for radial motion vanish for these spacetimes, 
they  should in some sense be looked  upon  as  ``minimally'' 
curved. They are free of Newtonian gravity. It is interesting to see their association with spacetimes 
describing  global  monopoles and textures. The  ansatz  could  be 
considered  as  a  prescription to  generate  ``minimally''  curved 
spacetimes that describe  topological defects, global monopoles and
textures. \\

\newpage
\n {\bf Acknowledgement :} KN thanks Jawaharlal Nehru Centre for
Advanced Research for an award of summer fellowship that facilitated
this work and IUCAA for hospitality. ND thanks Ramesh Tikekar for
a useful discussion on embedding.

\end{document}